\documentclass[12pt,a4paper]{article}
\usepackage{amsthm}
\usepackage[colorlinks,citecolor=blue,urlcolor=blue,filecolor=blue,backref=page]{hyperref}
\usepackage{graphicx}
\usepackage[utf8]{inputenc}
\usepackage{amsmath}
\usepackage{amsfonts}
\usepackage{tikz}
\usepackage{url}
\usetikzlibrary{shapes,arrows,positioning}

\begin{document}

%% *** Frontmatter *** 

%\begin{frontmatter}

\pagestyle{empty}
\let\svmaketitle\maketitle
\def\maketitle{\svmaketitle\thispagestyle{empty}}

\title{Structure learning of Bayesian networks involving cyclic structures}

%\title{\thanksref{T1}}
%\thankstext{T1}{<thanks text>}
%\runtitle{Structure learning of Bayesian networks involving cycles}

%\begin{aug}
\author{Witold Wiecek\footnote{Certara UK Ltd, 5th Floor Front, Audrey House, 16-20 Ely Place, London EC1N 6SN, United Kingdom},  Fr\'ed\'eric Y. Bois\footnote{Certara, Simcyp division, Level 2-Acero, 1 Concourse Way,Sheffield S1 2BJ, United Kingdom},  
%\author{\fnms{Sagnik} \snm{Datta}\thanksref{datta}}
Ghislaine Gayraud\footnote{Sorbonne Universit\'es, LMAC, Universit\'e de Technologie de Compi\`egne, France}}

\date{}
\maketitle

\begin{abstract}
Many biological networks include cyclic structures. In such cases, Bayesian networks (BNs), which must be acyclic, are not sound models for structure learning. Dynamic BNs can be used but require relatively large time series data. We discuss an~alternative model that embeds cyclic structures within acyclic BNs, allowing us to still use the factorisation property and informative priors on network structure. We present an~implementation in the linear Gaussian case, where cyclic structures are treated as multivariate nodes. We use a Markov Chain Monte Carlo algorithm for inference, allowing us to work with posterior distribution on the space of graphs. We provide a simulation study to evaluate the proposed model.
\end{abstract}


\section{INTRODUCTION}

Large-scale gene expression studies have invigorated interest in exploratory methods for evaluating patterns of association between random variables. The large number of random variables potentially considered and relatively small data sets challenge known structure learning approaches both conceptually and computationally. Graphical models are often used for to represent network structure and for statistical inference (see Lauritzen \cite{lauritzen_graphical_1996}). They depict the random variables of interest as nodes in a~graph, and conditional independence statements about them by presence or absence of graph edges. We focus on Bayesian networks (BNs), which represent probability distributions by means of directed acyclic graphs (DAGs) and are popular in learning structure of biological networks (e.g., Hausmeier, \cite{husmeier_reverse_2004}; Hausmeier and Werhli, \cite{husmeier_bayesian_2007}).  Under additional assumptions, described for example by Pearl \cite{pearl_causality:_2009}, directed edges of BNs can correspond to causal relationships between nodes. In BN models the joint probability distribution can be factorised between nodes and evaluated easily.

Yet, there are important cases in biology and other domains of study where we do want to consider cyclic structures such as feedback loops, which are common in gene transcription regulation networks and other biological networks (Alon,  \cite{alon_network_2007}).  In such cases DAGs cannot be used directly, as they do not offer a~sound representation of an~essential feature of the networks under analysis. Dynamic Bayesian networks (DBNs) offer an~alternative, by unrolling cycles, but can only be used when time variable is available. However, they multiply the number of nodes by the number of observation times and require dense and extensive data series, as discussed by  Ghahramani \cite{ghahramani_learning_1998}. 
%\textcolor{red}{Still, another problem with learning BNs with score-based methods is that graphs entailing the same conditional independence according to Pearl's \textit{d-separation} criterion \citep{pearl_causality:_2009} have the same likelihood \citep{geiger_parameter_2002}.}

We present a~different approach to modelling cyclic structures within a~graph. We contract such structures within the graph to derive an~associated acyclic graph. The contracted structures are treated as multidimensional random variables. For inference on graph structure, we use a~score-based implementation in the linear Gaussian case. Our approach is fully Bayesian, with scores being Bayesian marginal likelihoods. Our procedure uses the factorisation property of BNs and, to our knowledge, is novel. We implemented it in \textit{Graph\_sampler}, an~efficient C language software for simulated network generation and Bayesian inference on network structures. 
The informative priors we use, including on cyclic structures, imply that scores between two graphs may differ even if those graphs entail the same conditional independencies. In this sense our approach is an extension to previously proposed Bayesian approaches to network inference, e.g. by Mukherjee and Speed \cite{mukherjee_network_2008}, where the focus is on working with the full posterior distribution. 

This paper is organised in two parts. First, Section \ref{section:statistical_model} presents the statistical model that we use to score graphs, broken down between graph theory background, graph priors (including hyperparameters) and derivation of marginal likelihood. Then, Section \ref{section:examples} discusses choice of hyperparameters and examples of applications in graphs which involve cyclic structures, including computational benefit to Markov Chain Monte Carlo algorithms.

\section{STATISTICAL MODEL}\label{section:statistical_model}

Methods for learning structure of a~Bayesian network (presence or absence of edges between fixed nodes) can be categorised as either test-based methods for conditional independence or score-based methods. The latter tend to give more accurate results, according to o Acid et al., \cite{acid_comparison_2004}, and Cooper and Herskovits, \cite{cooper_bayesian_1992}, but their major disadvantage is the computational cost: since the number of possible graphs to consider grows super-exponentially with their number of nodes, exact inference on structure is a~hard problem. 

In our approach we use a~score-based method in a~Bayesian framework. 
That is, for any directed graph $\mathcal{G}$ (not necessarily acyclic) we define the graph's score $s(\mathcal{G}|D)$ conditionally on observed data $D$. The score is proportional to the marginal likelihood $s(D|\mathcal{G})$ and the prior distribution over the space of graphs $p(\mathcal{G})$, \textit{i.e.}, $s(\mathcal{G}|D) \propto s(D|\mathcal{G}) p(\mathcal{G})$. We define $s(\mathcal{G}|D)$ for any directed graph, not necessarily acyclic.

%The score is obtained from the graph's prior predictive density (p.p.d.), that is, 
To derive $s(D|\mathcal{G})$, we integrate over all of the model parameters as it is computationally more efficient than calculating the full posterior function and the parameters over which we integrate are not needed to make inferences about structure. We also make use of prior conjugacy, which helps quickly evaluate $s(D|\mathcal{G})$. A Metropolis-Hastings Markov Chain Monte Carlo (MCMC) algorithm is then used to sample graphs from their scoring distribution, \emph{cf.} Yu et al.  \cite{yu_advances_2004}, Zhou et al. \cite{zhou_bayesian_2004}, Datta et al.  \cite{datta_graph_sampler:_2017}. The result of inference is the distribution over space of graphs. For brevity, in the context of MCMC sampling we will refer to this result as a posterior distribution, even though some parameters have been integrated over.

As we will show later in this Section, graphs which imply same conditional independencies may have different marginal likelihoods. For acyclic graphs, this is due to use of informative priors. When cyclic structures are allowed, this difference may also arise by choice of hyperpriors, which can promote or penalise the occurence of contracted nodes. Before describing the statistical model, we will first introduce graph theory definitions on which the statistical model depends. The rest of this section will then describe the priors and data likelihood we use.

\subsection{Graph model}

In what follows, we assume that $\mathcal{G} = (V, A)$ is a~directed graph (with vertices set $V$ and directed edges set $A$), of a~given order $N = |V|$. We do not require for $\mathcal{G}$ to be acyclic, but edges from a~node to itself (\textit{auto-cycles}) are not allowed. We use terms \textit{graph} and \textit{network} interchangeably. A \textit{walk} is a sequence of $k$ directed edges $(v_i, u_i)$ such that $u_i = v_{i+1}$ for $i=1,2,\ldots,k-1$. A \textit{path} is a walk where all vertices ($\lbrace v_1, v_2, \ldots, v_k, u_k \rbrace$) are unique.

We say that a~graph is \textit{strongly connected} if for every pair of vertices there exist paths in each direction between the two. A \textit{strongly connected component} (SCC) of a~graph is a~maximal subgraph that is strongly connected. By definition, every cycle is a~strongly connected (although not maximal) subgraph. Not all SCCs are cycles, however; \textit{e.g.} a~``flat eight'' graph of three nodes $A\rightarrow B\rightarrow C\rightarrow B \rightarrow A$ is strongly connected but $B$ is traversed twice to get from $A$ to $C$, hence it is not a~cycle. 
We call single node components \textit{ordinary}. When referring to SCCs we typically mean non-ordinary SCCs (SCCs of more than one node), unless explicitly stated. For each graph we can create a~partition of its vertices into a set of strongly connected components. We denote such partition by $\mathrm{SCC}(\mathcal{G})$. It can be performed in linear time, as first proposed by Tarjan \cite{tarjan_depth-first_1972}. Since most of the strongly connected components which we encounter in structure learning of biological networks are graph cycles, we will also interchangeably use the term ``cyclic structures'' throughout the paper.

\begin{figure}[!ht]
\centering
\includegraphics[width=1\textwidth]{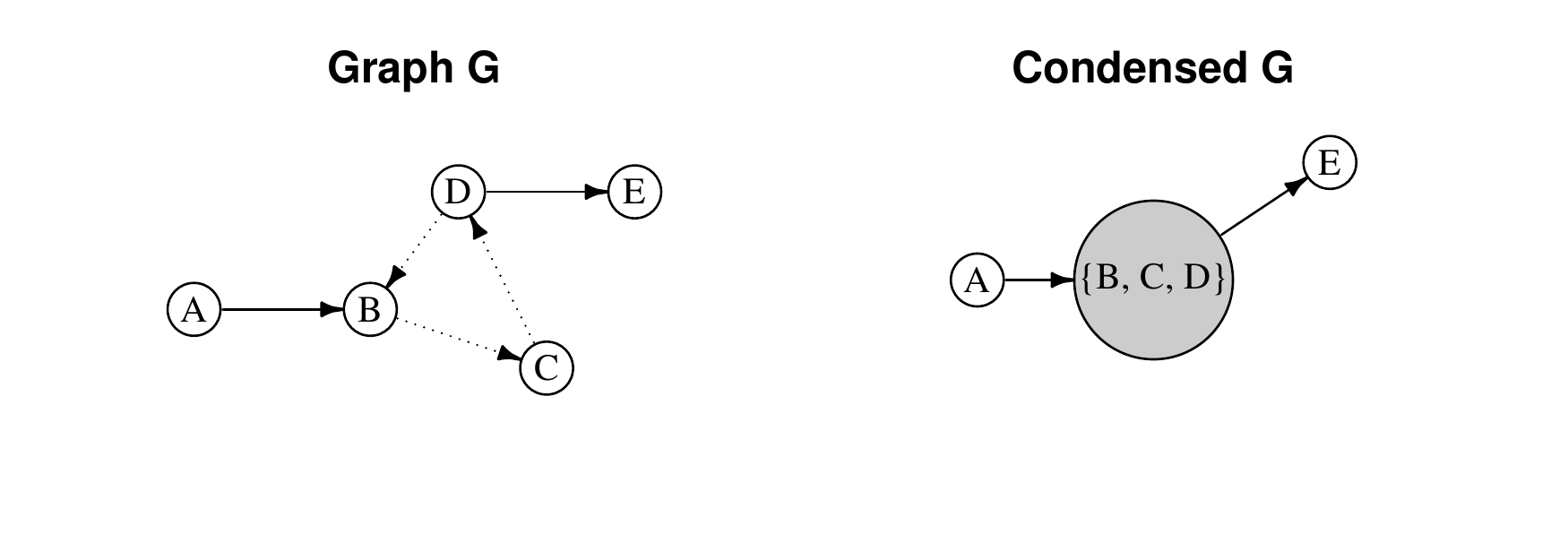}
\caption{Graph $\mathcal{G}$ with 5 nodes where nodes B, C and D form a~cycle and are contracted to a~single node after condensing the graph. Here $\mathrm{SCC}(\mathcal{G})=\{A, \{B,C,D\},E \}$}
\label{fig:condensed}
\end{figure}

For any directed graph $\mathcal{G}$, we can create an~associated \textit{condensed graph}, $\mathcal{G}_c$, by repeatedly \textit{contracting} edges (replacing a pair of vertices connected by an edge by a single vertex, retaining all directed edges) in each strongly connected component until each component corresponds to a~single vertex. By construction such graph is acyclic. An illustration is provided in Figure~\ref{fig:condensed}.

A \textit{Markov equivalence class} is a set of DAGs that have the same skeleton (set of edges without regards to direction) and \textit{v-structures} (sets consisting of a~child and its two parents that are not themselves connected). Members of the equivalence class encode same conditional independencies, as shown by Verma and Pearl, \cite{verma_algorithm_1992}, in the context of causal inference. Various algorithms have been proposed to learn Markov equivalence classes of causal graphs, \textit{e.g.} by Chickering \cite{chickering_optimal_2003}. 
When cyclic structures are present and we are working with condensed graphs, we assume that the conditional independencies implied by the Bayesian network are only the ones that are implied by the condensed graph $\mathcal{G}_C$. As $\mathcal{G}_C$ is a DAG, we can take advantage of the Markov property and factorise the joint probability distribution over nodes of the condensed graph $\mathcal{G}_c$. We then model the initial $\mathcal{G}$ by considering multivariate distributions on each of the strongly connected components.

\subsection{Priors on graph structure}

All possible graphs of a~certain size are not equally plausible \textit{a priori} and we should consider that prior knowledge on graph structure in our inferences. Distributions and parametrisations of the following priors have been previously described by Mukherjee and Speed \cite{mukherjee_network_2008} and by Datta et al. \cite{datta_graph_sampler:_2017}: 

\begin{itemize}
\item \textit{Bernoulli priors} on existence of individual (directed) edges, specified by providing a~square matrix of edge probabilities. Gene
association studies can provide this type of prior knowledge.
\item \textit{Concordance prior} between the graph adjacency matrix and an~edge requirement matrix, where each edge is classified as desired, not desired or no preference. This penalizes networks too different from a canonical one (although, tuning this pseudo-prior is not very easy).
\item \textit{Degree prior} on the distribution of node degrees $d(v)$ in the graph, using a~power law with parameter $\gamma$. The degree distribution of many
physical networks appear to follow approximately such a power law ( Barab\`asi and Albert \cite{barabasi_emergence_1999}).
\item \textit{Edge count prior} on the expected graph size.
\item \textit{Motif prior} on the count of triangular feed forward and feedback loops in the network, as discussed in Bois and Gayraud \cite{bois_probabilistic_2015}.
\end{itemize}

All the above priors are specified on the input graph $\mathcal{G}$ (and not $\mathcal{G}_c$). To work with cyclic structures, we introduce two additional structure priors on strongly connected components:

\begin{itemize}
\item \textit{Prior on number of strongly connected components.} Consider partition of graph $\mathcal{G}$ into $\mathrm{SCC}(\mathcal{G})$, with ordinary nodes discarded. We define a~prior on the number of strongly connected components (of at least two nodes) $|SCC(\mathcal{G})|$  via a~Poisson distribution: $
|\mathrm{SCC}(\mathcal{G})| \sim  \text{Poisson}(\lambda_{\mathrm{SCC}})$, with $\lambda_{\mathrm{SCC}}>0$.

\item \textit{Prior on the size of strongly connected components.} We also define a~prior $p_{\mathrm{SCC}}$ on the size of all components present in $\mathcal{G}$, using a~power law: 
$$p_{\mathrm{SCC}} \propto \prod\limits_{S \in SCC(\mathcal{G})} (|S|)^{-\gamma_{\mathrm{SCC}}},$$ with $\gamma_{\mathrm{SCC}}>0$.
\end{itemize}

\noindent The global prior probability of graph, $p(\mathcal{G})$, is then proportional to a~product of all the priors specified.

\subsection{Data likelihood}\label{section:data_likelihood}

% https://pdfs.semanticscholar.org/d286/2c3ab86bce3d0c11f8a617d1781cc45aeffa.pdf

% Decompose p(G|D) = c p(D|G)p(G)
% Name local parameters, say they are local, show the integration operation DIRECTLY in section 2 (as Heckerman does)
% Mention "global parameter independence" and/or quote 3 assumptions

% Re-do notation more in terms of p(x)
% Q for FB: what is the best multivariate regression reference that I can quote here?

% We now turn to calculating the marginal likelihood of data, given any directed graph $\mathcal{G}$, including graphs with cycles. 
%Our ultimate aim is to fall back on the key feature of Bayesian networks: the factorisation of its associated probability distribution. We achieve that by condensing $\mathcal{G}$ into the acyclic $\mathcal{G}_c$.
Let $D = \mathbf{x} = (x_1,...,x_N)$ denote the observed data on $N$ nodes, where $x_i$ is an~\textit{n}-dimensional vector, with $n$ the number of data points per node. When we condense $\mathcal{G}$, we bifurcate its nodes (and corresponding $x$'s) into nodes obtained by contraction (corresponding to strongly connected components of at least two nodes) and ordinary nodes (single-node SCCs), for which no contraction was needed. Thus we represent $x$ as:

\begin{eqnarray}
\mathbf{x} = ({x_1}^D, {x_2}^D, ..., {x_{N_1}}^D, \mathbf{{x_1}^L, {x_2}^L, ..., {x_{N_2}}^L}) \nonumber
\end{eqnarray}

\noindent where $N_1$ denotes the number of ordinary nodes and thus various $x_{i}^{D}$'s are just relabelled $x_i$'s from the original data set; $N_2$ represents the number of non-ordinary SCCs. For $j = 1, 2, ..., N_2$, $\mathbf{{x_j}^L} = \mathbf{(x_{j_1},...,x_{j_m})}$ is a~data set of the $m$ node members of the $j$-th non-ordinary SCC. Given this partitioning of data, for any graph $\mathcal{G}$ the likelihood can be factorised into a~product over ordinary and non-ordinary components:

\begin{eqnarray*}
s(D \vert \mathcal{G}) = \prod\limits_{i=1}^{N_1} s({x_i}^D \vert Pa({x_i}^D)) . \prod\limits_{j=1}^{N_2} s(\mathbf{{x_j}^L} \vert Pa(\mathbf{{x_j}^L})) 
\end{eqnarray*}

\noindent where $Pa(\cdot)$ denotes the parent nodes of $(\cdot)$ in $\mathcal{G}_c$. If there are no parents, $s(x_i \vert \emptyset) = s(x_i)$. The remainder of this section describes how to obtain the terms $s({x_i}^D \vert Pa({x_i}^D))$ and $s(\mathbf{{x_j}^L} \vert Pa(\mathbf{{x_j}^L}))$ under a linear Gaussian model.

\subsubsection{Marginal likelihood for contracted nodes}

Let us consider a~strongly connected component of $m$ nodes $j_1, j_2, j_3,\ldots, j_m$. As defined above, $Pa(\mathbf{{x_j}^L})$ is the set of its parents in the condensed graph $\mathcal{G}_c$, that is, 
$\cup_{i = 1, 2,\ldots, m}{Pa(x_{j_i})}$.
% the union set of parents of each of the $m$ nodes.

We model the distribution of $\mathbf{{x_j}^L}\vert Pa(\mathbf{{x_j}^L})$ using a~linear multivariate Gaussian model; setting $Y = \mathbf{{x_j}^L}$, the model can be expressed as

\begin{equation}
Y = X \theta + \epsilon,
\label{reg2}
\end{equation}

\noindent where $Y$ is a~matrix of dimension $n \times m$, $X$ is the design matrix of size $n \times k$ with ones in its first column and $Pa(\mathbf{{x_j}^L})$ in the remaining columns, so that $k$ = $\dim(Pa(\mathbf{{x_j}^L}))+1$. 
Coefficient matrix $\theta$ is of dimension $k \times m$, while $\epsilon$ is a~$n \times m$ dimensional matrix, $\epsilon=(\epsilon_1, \epsilon_2,.., \epsilon_n)$, where all $\epsilon_i$'s are independent and identically distributed with a multivariate Gaussian distribution $\mathcal{N}_m (0, \Sigma)$. 

\noindent Under this model, the likelihood $L$ is multivariate normal and can be expressed as:

\begin{align}\label{reg3}
L(Y, X, \theta, \Sigma) =& \dfrac{1}{(2 \pi)^{nm/2}} \vert \Sigma \vert^{-n/2} \\ 
      &\exp \left\lbrace\dfrac{-tr[\Sigma^{-1} (Y - X \theta)^{t} (Y - X \theta)]}
       {2} \right\rbrace,\nonumber
\end{align}

\noindent where $tr(\cdot)$ denotes the trace of $(\cdot)$. 

For $\theta$ and $\Sigma$, we consider independent priors, \textit{i.e.}, $p(\theta, \Sigma)= p(\theta) p(\Sigma)$. 
In order to have an~analytically explicit form of the marginal likelihood we define an~improper (locally uniform) prior on $\theta$, $p(\theta) \propto \textrm{constant}$. For the prior distribution of $\Sigma$, we use an~$m$-dimensional inverse Wishart distribution, denoted $\mathcal{W}_m^{-1}(\kappa, q)$:

\begin{equation}\nonumber
p(\Sigma) = \dfrac{\vert \kappa \vert ^{\frac{q}{2}} \vert \Sigma \vert ^{\frac{-(q+m+1)}{2}}}{2 ^{\frac{mq}{2}} \Gamma_{m}(\frac{q}{2})} \exp \left\lbrace - \dfrac{1}{2} tr (\kappa \Sigma ^{-1}) \right\rbrace,
\end{equation}

\noindent  where $\kappa$ is a~positive definite scale matrix and scalar $q$ (degrees of freedom) is strictly positive; $\Gamma_{m}$ is the multivariate gamma function. We refer to this prior as \textit{constant-Wishart}, to distinguish it from other possible models outlined below. 

% The joint posterior distribution is 
% 
% \begin{align}
% P(\theta, \Sigma \vert Y,X) \propto 
% & \ C_0 \vert \Sigma \vert^{-n/2} \vert \Sigma \vert ^{\frac{-(q+m+1)}{2}}  \nonumber \\
% & \ \exp \left\lbrace - \dfrac{tr [\Sigma^{-1}(Y - X \theta)^{t} (Y - X \theta)] + tr (\kappa \Sigma ^{-1})}{2}\right\rbrace
% \label{eq3}
% \end{align}
% 
% \noindent where $C_0$ is given by:
% 
% \begin{equation}
% C_0 = \dfrac{\vert \kappa \vert ^{\frac{q}{2}}}{2 ^{\frac{mq}{2}} \Gamma_{m}(\frac{q}{2})} \dfrac{1}{(2 \pi)^{nm/2}}. \nonumber
% \end{equation}

The least square estimate $\hat{\theta}$ for the matrix $\theta$, and the sample co-variance matrix $A_0$ are given by:

\begin{eqnarray}
\hat{\theta} &=& (X^t X)^{-1}X^t Y    \nonumber\\
A_0 &=& (Y - X \hat{\theta})^t (Y- X\hat{\theta})
\label{eq4}
\end{eqnarray}

\noindent where $\hat{\theta_i}$ is the least square estimate for the $i$-th column of $\theta$. To define $\hat{\theta}$, we need $ (X^t X)^{-1}$ to exist and thus have a constraint $k \leq n$. 

% From equations (\ref{eq3}) and (\ref{eq4}) the joint posterior distribution of both $(\theta, \Sigma)$ can be expressed as:
% \begin{equation}
% P(\theta, \Sigma \vert Y) \propto C_0  \vert \Sigma \vert^{-n/2} \vert \Sigma \vert ^{\frac{-(q+m+1)}{2}} \exp \left\lbrace\dfrac{-1}{2} tr\Sigma^{-1}[A_0 + (\theta - \hat{\theta})^t X^t X(\theta - \hat{\theta})]\right\rbrace \exp \left\lbrace - \dfrac{1}{2} tr (\kappa \Sigma ^{-1}) \right\rbrace
% \end{equation} 
% which can further be simplified and expressed as:

The joint posterior distribution of $(\theta, \Sigma,Y)$ conditional on $X$ can be expressed as

\begin{equation}
\label{model_posterior} 
P(\theta, \Sigma, Y \vert  X) =   P(\theta \vert \Sigma, Y,X) \cdot P(\Sigma \vert Y,X) \cdot  P(Y\vert X), 
\end{equation}

\noindent where  $P(Y\vert X)$,  $P(\theta \vert \Sigma, Y,X)$ and  $P(\Sigma \vert Y,X)$  are of the form:

\begin{align}\label{pre} 
P(Y\vert X)  = & (2 \pi)^{\frac{m(k-n)}{2}} 2^{\frac{m(p-q)}{2}} \dfrac{\Gamma_{m}(\frac{p}{2}) }{ \Gamma_{m}(\frac{q}{2})} 
\vert\kappa \vert ^{\frac{q}{2}}  \vert X^t X \vert ^{-\frac{m}{2}} \vert \kappa + A_0 \vert ^{-\frac{p}{2}},\\
% C_0 &=& \dfrac{\vert \kappa \vert ^{\frac{q}{2}}}{2 ^{\frac{mq}{2}} \Gamma_{m}(\frac{q}{2})} \dfrac{1}{(2 \pi)^{nm/2}} (2 \pi)^{mk/2} \vert X^t X \vert ^{\frac{-m}{2}} \vert \kappa + A_0 \vert ^{-\frac{p}{2}} 2^{\frac{mp}{2}} \Gamma_{m}(\frac{p}{2}), \label{pre}\\
P(\theta \vert \Sigma, Y,X) = & \dfrac{\vert X^t X \vert ^{\frac{m}{2}} \vert \Sigma \vert^{-k/2}}{(2\pi) ^{\frac{mk}{2}}}
\exp \left\lbrace -\dfrac{1}{2} tr\Sigma^{-1} (\theta - \hat{\theta})^t X^t X(\theta - \hat{\theta})\right\rbrace, \nonumber\\
P(\Sigma \vert Y,X) = & \dfrac{\vert \Sigma \vert^{-(p+m+1)/2}}{2 ^{\frac{mp}{2}}\Gamma_{m}(\frac{p}{2})}
\exp \left\lbrace - \dfrac{1}{2} tr \Sigma ^{-1}(\kappa + A_0) \right\rbrace \vert \kappa + A_0 \vert ^{\frac{p}{2}},\nonumber
\end{align}

\noindent with $p = q+n-k$. 

\noindent When $\theta$ and $\Sigma$ are integrated out from (\ref{model_posterior}), we obtain (\ref{pre}), which corresponds to the marginal likelihood function of the strongly connected component under the \textit{constant-Wishart} model, that is, 

$$P(Y\vert X)= s(\mathbf{{x_j}^L}\vert Pa(\mathbf{{x_j}^L})).$$

\subsubsection{Marginal likelihood for ordinary nodes}

In the case of an~uncondensed acyclic graph, possible forms of marginal likelihood $s({x_i}^D \vert Pa(x_i^D))$ have been discussed previously by Datta et al. \cite{datta_graph_sampler:_2017}. With a~univariate linear regression model on $x_i$'s parents, using a~classical Normal-Gamma conjugate prior (inverse Gamma on the scale and conditional normal on the mean), integrating out these parameters leads to a~multivariate Student's $t$ distribution. Zellner and Dirichlet  likelihoods are other possible choices and also described therein, together with the choice of likelihood parameters' hyperpriors.

In the case where cyclic structures are allowed, we treat the ordinary nodes as 1-dimensional special cases of the constant-Wishart prior, owing to the fact that the inverse Wishart distribution with parameters $q, \kappa$ is the multivariate version of the inverse Gamma distribution with parameters $(q/2, \kappa/2)$. We show the equality of the two marginal likelihoods in Supplement S1. 
%\ref{supplement:equivalence}.

\section{MODEL PROPERTIES AND APPLICATIONS}\label{section:examples}

In this section we will discuss three topics: how the choice of hyperparameters impacts the graph score in SCC cases; how inference on structure is accomplished with an MCMC algorithm; some examples of applications of our approach. The examples will illustrate role that priors and SCCs play in both learning network structure and the computational aspect of inference.
%We first present a computational example for how allowing SCCs in MCMC procedures when convergence to a~DAG would be difficult or impossible. We follow with two examples of inference on graph structure under our data generating model, \textit{i.e.} linear relationships between nodes and additive normal noise. 
%First is a~simple example of a~seven-node graph. We show how the choice of prior is necessary learning the correct graph model. In the second example, we illustrate the performance of the approach on 50-node graphs, with a~special focus on the role that inverse Wishart hyperparameters play in detecting SCCs. 

\subsection{Likelihood equivalence in constant-Wishart case}

Geiger and Heckerman, \cite{geiger_parameter_2002}, discuss conditions under which graphs in an~equivalence class will have the same likelihood. A Gaussian model with inverse Wishart prior is such a~case. This notion of equivalence can be extended to marginal likelihoods. Heckerman et al.,  \cite{heckerman_learning_1995}, present an~additional assumption sufficient for marginal likelihood equivalence. It requires that the Jacobian of the one-to-one mapping between two parameters sets associated with two distribution equivalent graphs exists and the priors of the two parameters sets must be equal after applying the change of variables formula.

However, in our case this property no longer holds since we consider independent prior on each parameter set attached to a~single term in the likelihood factorisation. For the case of a~two-node graph, we derive the marginal likelihood explicitly in the Supplement S2
%\ref{supplement:loop_dag} 
and show how $s(D \vert A \rightarrow B) \neq s(D | B \rightarrow A)$. In the case of Markov-equivalent DAGs with equal priors, the differences in graph score are due to sampling variance and tend to 0 with increasing $n$. However, in the case of SCCs, the scores will differ between equivalent DAGs due to marginalisation of the likelihood and the difference doesn't tend to $0$ with growing sample, but rather depends on the choice of hyperparameters of the inverse Wishart prior distribution of $\Sigma$.

Hyperparameters under the constant-Wishart model are the scale matrix $\kappa$ and the scalar $q$ (degree of freedom). As mentioned, we use inverse Wishart prior as it is conjugate and allows us to obtain the marginal likelihood analytically. Additionally, this prior, when informative, can be interpreted in terms of equivalent sample size. If $X \sim \mathcal{W}^{-1}(\kappa, q)$ then

\begin{eqnarray}
\mathrm{E}(X_{ij}) &=& 
   \frac{\kappa_{ij}}{q - m - 1}, 
   \quad \text{when} \ q > m+1, \label{equation:expected_value} \\
\mathrm{Var}(X_{ij}) &=& 
   \frac{(q-m+1)\kappa_{ij}^2 + (q-m-1)\kappa_{ii}\kappa_{jj}}{(q-m)(q-m-1)^2(q-m-3)}, 
   \quad \text{when} \ q > m+3. \nonumber 
\end{eqnarray}

However, there are known issues with using inverse Wishart priors: they imply relationships between variances and covariances and use a~single parameter ($q$) to describe precision on all parameters. When $q > 1$, the prior may be biased when the true variance is low, even with large sample sizes, as discussed by Gelman  \cite{gelman_prior_2006}; see also Alvarez et al.,  \cite{alvarez_bayesian_2014}, for a~simulation study.

In inference on variance-covariance matrices, it is typical to assume $\kappa = I_m$ (identity matrix of order $m$) and $q = m+1$. Since $q$ is responsible for the precision of the prior and can be interpreted in terms of sample size equivalence, setting a~low $q$ is a~good default choice. In such case, prior marginal distributions of correlations are uniform on $(-1,1)$. However, this goes against our intuition: typically, we assume \textit{a priori} that nodes of an SCCs are going to be strongly correlated; exactly how strongly depends on context and objectives of analysis. 
%We explore the impact of prior on SCC scores in detail in supplement \ref{supplement:scc_likelihood}. 
By default, we propose to set $q=m+1$ and $\kappa$ to 1 on diagonal elements and to 0.5 on off-diagonal. This creates a monotonic prior on correlation and ensures higher marginal likelihood for SCC than all DAG graphs when the true correlation is higher than 85\%-90\%. This choice is explored and explained below.

Let us define $d$ as the difference in log marginal likelihoods between a graph $\mathcal{G}_{\mathrm{SCC}}$, where all $m$ nodes form an~SCC, and a complete DAG $\mathcal{G}_{\mathrm{DAG}}$, that is, a DAG with no missing arcs (no conditional independencies):
$$d = \log{s(D \vert \mathcal{G}_{\mathrm{SCC}})} - \log{s(D \vert \mathcal{G}_{\mathrm{DAG}})},$$
conditional on the same data $D$. Positive $d$'s indicate that the SCC is more likely than the DAG.

We now briefly explore the behaviour of $d$ in SCCs of different sizes by means of simulated data and show that it is predictable in ways that may be useful in practical applications. For all the examples presented in this section we generated $n = 10,000$ draws of data of $m$ nodes from multivariate normal distribution with means 0 with each node having fixed variance $\sigma^2$ and same correlation $\rho$ with all other nodes. We started with $\sigma^2 = 1$ and varied $\rho$ between 0 and 1. We compared the SCC against one DAG only because all complete DAGs form an equivalence class.

% As mentioned, it is typical to set $q = m+1$ and $\kappa = I_m$. 
% However, even with the default value $q = m + 1$, $\kappa$ has an~impact on choice between SCCs and DAGs. Setting $\kappa = I_m$ in this situation leads to uniform priors on correlation. This can be desirable in inference on unknown variance-covariance matrix, but our goal in this case is to distinguish cyclic structures from DAG parts of the network: therefore a~sensible choice would be to have a~prior that ``favours'' loops when correlation is higher. 

\begin{figure}[!ht]
\centering
\includegraphics[width=1\textwidth]{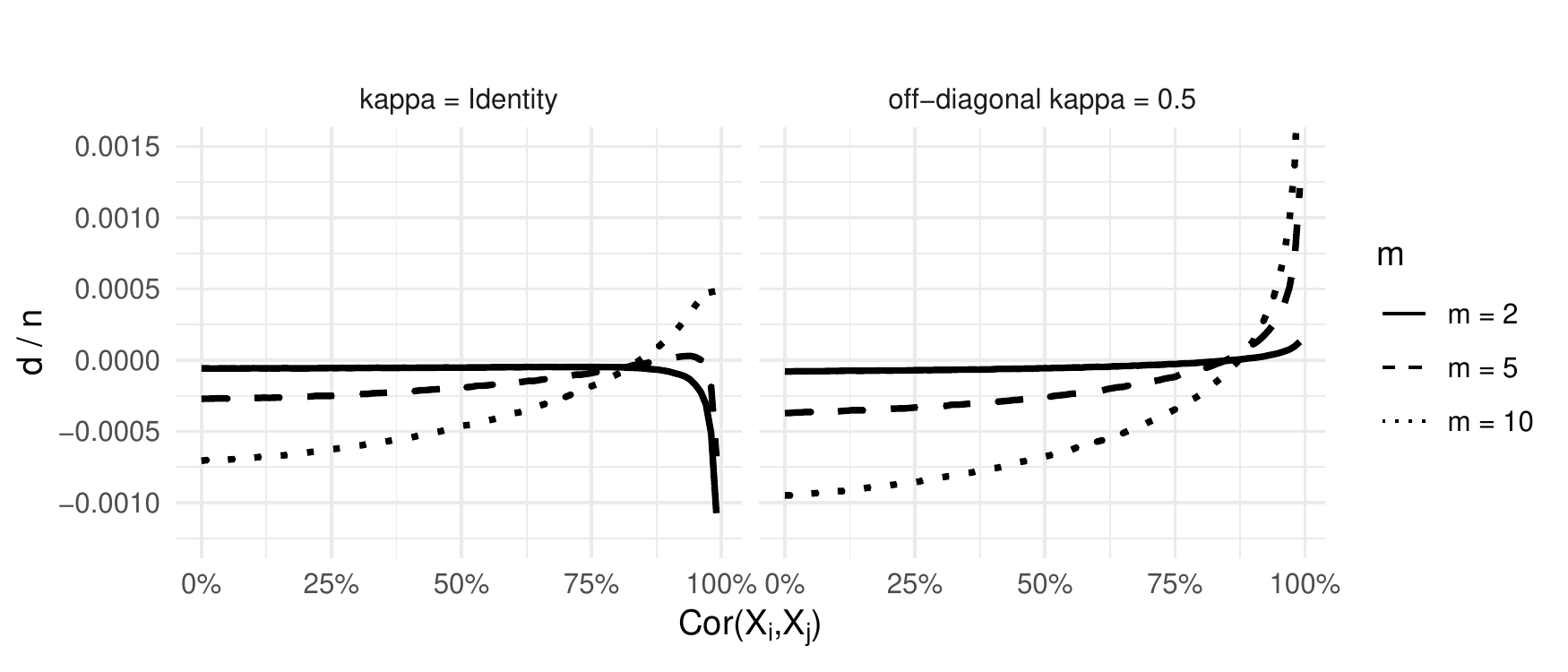}
\caption{Difference $d$ between SCC and DAG graph scores (divided by $n$ data) as a~function of true correlation between pairs of nodes (x axis) and number of nodes $m$. The hyperparameters are set to $q = m+1$ and $\kappa = I_m$ in the left panel. In the right panel we change $\kappa_{ij} = 0.5$ for $i \neq j$.}
\label{figure:lvg}
% lvg = 'loop versus DAG'
\end{figure}

We illustrate behaviour of $d$ as a function of the off-diagonal elements of $\kappa$ in two panels of Figure \ref{figure:lvg}. With $\kappa = I_m$ the sign of $d$ is not consistent, but when the off-diagonal elements of $\kappa$ are set to $0.5$ everywhere, $d$ is positive when the true correlation in data exceeds 85\%-90\% threshold. Therefore we use $0.5$ as the default choice of prior. Difference grows larger as $m$ increases.

The behaviour of $d$ is also sensitive to variance of random variables. %This is illustrated in Figure \ref{figure:lvg2}: 
If $\kappa$ is misspecified, the SCCs are always preferred for low variances and DAGs are always preferred for high variances. However, standardising the inputs can solve this problem, as will scaling $\kappa$ by sampling variances of each node. This can be done automatically in software implementations and in both cases will ``bring back'' the behaviour of $d$ to exactly what is seen in Figure \ref{figure:lvg}.

\begin{figure}[!ht]
\centering
\includegraphics[width=1\textwidth]{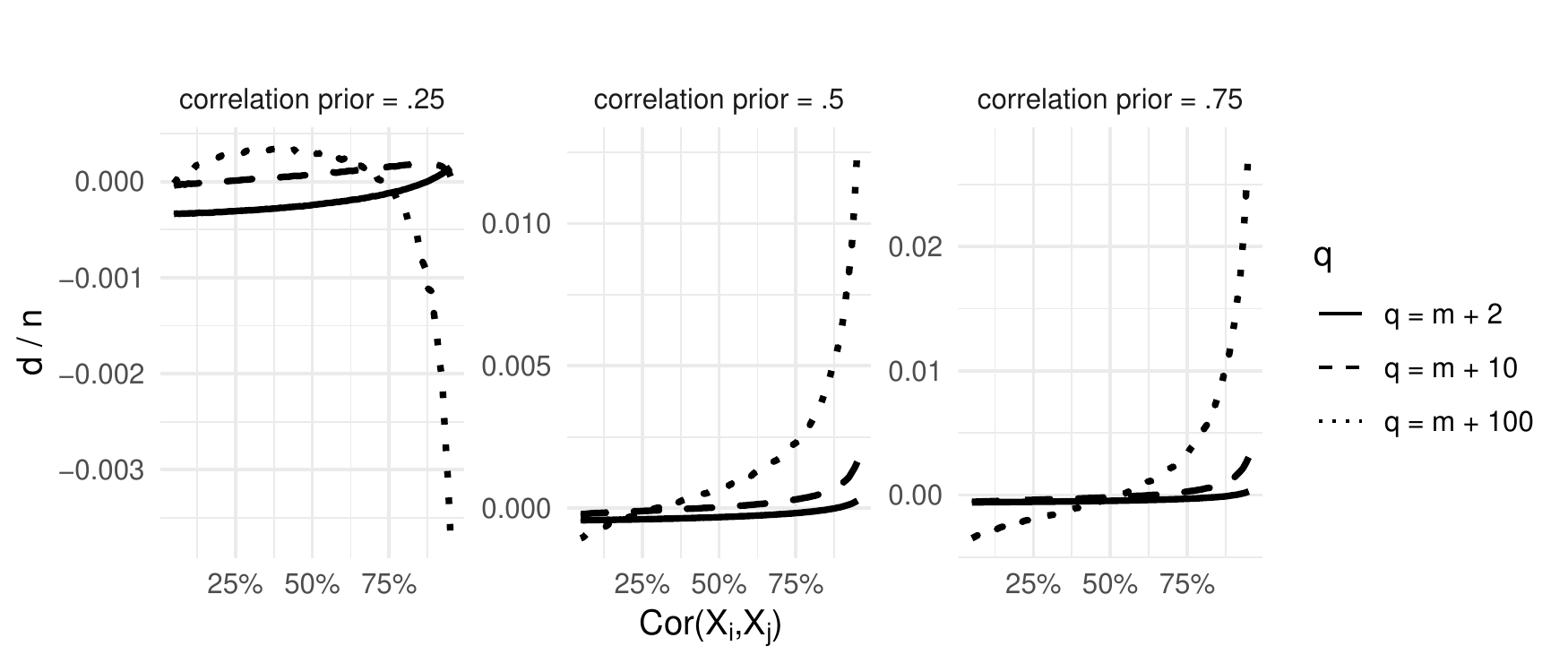}
\caption{Difference $d$ between SCC and DAG graph scores (divided by $n$ data) under informative priors on correlation. $\kappa$ is scaled according to $q$ to yield desired mean correlation: 0.25 in the left panel, 0.5 in the middle, 0.75 in the right panel.}
\label{figure:lvg3}
\end{figure}

As indicated by (\ref{equation:expected_value}), we can put a~prior on correlation between two elements to any mean $\rho$ by setting off-diagonal elements of $\kappa$ to $(q-m-1)\rho$, and to any variance by adjusting $q$. In practice, we can use this to manipulate the sign of $d$, thus allowing us to choose the level of correlation at which SCCs will be chosen over DAGs different from the 85\%-90\% threshold. This is illustrated in Figure \ref{figure:lvg3}.

%\begin{figure}[ph]
%\centering
%\includegraphics[width=1\textwidth]{figures/fig_loop_vs_dag2.pdf}
%\caption{Difference $d$ between SCC and DAG graph scores (divided by $n$ data) when %prior is wrongly specified with regards to true data generating mechanism (variance %of 100 in the left panel or .01 in the right panel). In both cases the prior is $q %= m+1$ and $\kappa_{ij} = 0.5$ for $i \neq j$ or 1 when $i=j$.}
%\label{figure:lvg2}
%\end{figure}

\subsection{MCMC algorithm for inference}

The computer code needed to perform all of the examples has been implemented in the latest version of the \textit{graph\_sampler} software for MCMC inference on graphs, previously introduced by Bois and Gayraud \cite{bois_probabilistic_2015}. 

Written in ANSI-standard C language, the full software is freely available at \textit{www.nongnu.org/graphsampler} under the terms and conditions of the GNU General Public License, as published by the Free Software Foundation. 

\textit{Graph\_sampler} uses Metropolis-Hastings algorithm to sample graphs from a~scoring distribution. The proposals in the algorithm are edge additions or deletions, drawn according to a Bernoulli prior on the graph edges. For DAGs, the score of the proposal is then evaluated by calculating the difference in scores on the child node in the proposed addition or deletion. In all cases convergence to the target distribution can be checked by calculating the Gelman-Rubin statistic (Gelman and Rubin, \cite{gelman_inference_1992}) on chains of graph adjacency matrices. Convergence check function is included as part of the software.

When cyclic structures are allowed, the algorithm is modified to take into account situations where the condensed graph changes, (\textit{i.e.} SCCs are created or deleted). Multiple nodes are affected in such situations and need to have their scores recalculated. We devised an additional decision rule to only condense graph (using Tarjan's algorithm) when necessary and recalculate likelihood on the minimal set of nodes that may be affected by additions and deletions. It is presented in Supplement S3.
%\ref{supplement:flowchart}.

The MCMC approach yields a set of $n$ graphs sampled from the posterior distribution. We represent them by their adjacency matrices $A^{(1)}, \ldots, A^{(n)}$. Such a sample can be used to approximate posterior probabilities of occurence of edges or motifs. For example, the probability of an~edge from $i$ to $j$, $p_{ij}$, is obtained by calculating $\hat{p_{ij}} = \sum{A}_{ij} / n$. However, such probabilities have to be treated with caution when cyclic structures are allowed. Depending on the objectives of analysis, we can either be interested in $\hat{p_{ij}}$ defined as above or the probability of $i$ and $j$ being part of the same SCC ($p^{SCC}_{ij}$) or of $i$ being parent of $j$, but not in the same SCC ($p^{DAG}_{ij}$).

\subsection{MCMC convergence in SCC setting}
%This is case_conv/ subfolder in the project

If the MCMC algorithm for graph inference operates only by adding or removing edges at each step, reversing the direction of an~existing edge can be difficult. It requires two operations: a~deletion followed by an~addition. The first step will often (\textit{e.g.}, in situations where two nodes are highly correlated) have an~extremely low probability. Using tempered MCMC methods can solve this problem (see Baker  et al., \cite{barker_mc4:_2010}) but requires fine tuning of the tempering algorithm. Using SCCs provides a~simpler solution: an~addition (creating an~SCC) followed by a~deletion. Thus for some problems, allowing cyclic structures can be helpful even if we know that  the true network is acyclic as it can avoid traversing these ``probability wells''.

\begin{figure}[!ht]
\centering
\includegraphics[width=1\textwidth]{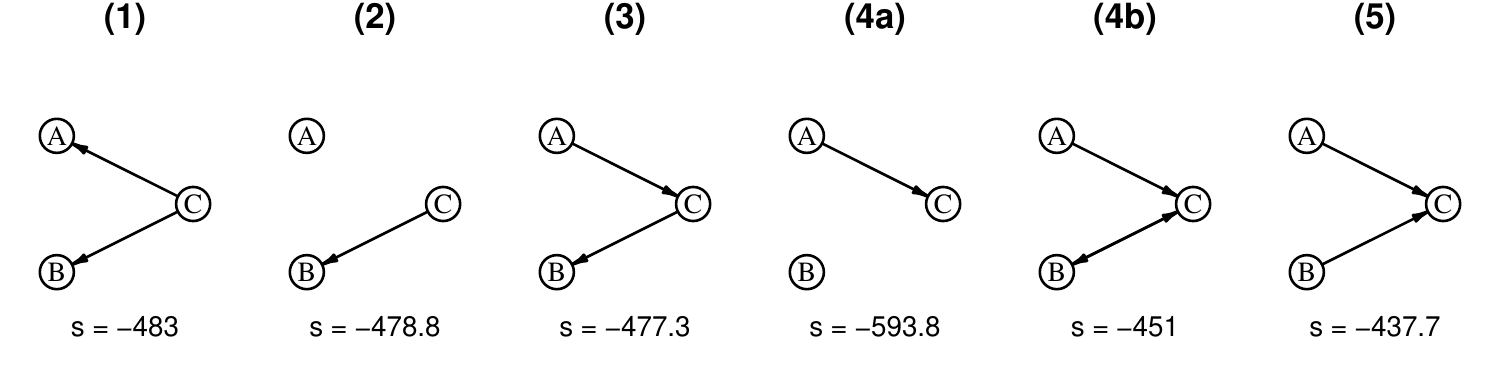}
\caption{"Inverting" a v-structure in five steps. For each graph value $s$ is $s(D \vert \mathcal{G})$. Leftmost graph is the starting graph for MCMC and the true generative mechanism is the rightmost. While deleting some edges such as $CA$ can be ``easy'' in terms of log likelihood difference (steps 2 and 3),  deletion of the $CB$ edge (step 4a) is difficult as it leads into a ``probability well''. Adding the edge $BC$ (4b), which results in an SCC, offers a way to reach (5) while avoiding the well.}
\label{fig:conv}
\end{figure}

We illustrate this with a~simple example of "inverting" a~v-structure. Assume $X_A \sim \mathcal{N}(0, 1)$, $X_B \sim \mathcal{N}(0,1)$ and $X_C = X_A - 5X_B + \epsilon$ with $\epsilon \sim \mathcal{N}(0,1)$. We draw 100 realisations of each random variable. 
Assume the MCMC sampler starts from a~graph model $A \leftarrow C \rightarrow B$ with score of $-472.5$. Assuming that we are working with DAGs only, any path to the true generating graph requires removal of $CB$ edge. This is shown in Figure \ref{fig:conv}.

%\begin{table}[!ht]
%\caption{Edge probabilities (rows: from, columns: to) obtained with DAGs only (left) and when %inference allows SCCs (right). The true generative model is $A \rightarrow C, B \rightarrow %C$. The probability of (wrong) edge $CB$ stays at 1 in first case. The probability of the %``reversed'' edge $BC$ is close to one after allowing SCCs.}
%\centering
%\begin{tabular}{l ccc | l ccc}
%
%\multicolumn{4}{c}{Case 1: only DAGs} & \multicolumn{4}{c}{Case 2: SCCs allowed} \\ 
%  &     A &     B &     C &   &     A &     B &     C \\ \hline
%A & 0.000	& 1.000	& 0.820 & A & 0.000	& 0.202 &	0.960 \\
%B & 0.000	& 0.000	& 0.000 & B & 0.270	& 0.000 &	0.977 \\
%C & 0.027	& 1.000	& 0.000 & C & 0.109	& 0.000 &	0.000 \\
%\end{tabular}
%\label{tab:convergence}
%\end{table}

The ``well'' is a~score difference of around 100 (therefore on average we would need $e^{100}$ Metropolis-Hastings proposals to remove $CB$). Using SCCs easily circumvents this by creating an SCC involving the $B \leftrightarrow C$ SCC before deleting $CB$. 
%The resulting edge probabilities from both models are presented in Table \ref{tab:convergence}.

\subsection{Linear model with additive noise}

The following example is straightforward, but difficult to correctly estimate. For this, we slightly expanded the graph from Figure~\ref{fig:condensed} by adding node $P$, a parent to $A$,  and $Q$, a child to $E$. We assumed a linear relationship and generated 100 draws for each node as follows: for $j$-th node, $i$-th generated value $x_j(i) = Pa_{x_j}(i) + \epsilon_{ij}$, where $Pa_{x_j}$ are data for the parent node of $X_j$ (for node $P$ we set the mean to zero) and $\epsilon_{ij}$ are i.i.d. with $\mathcal{N}(0, 5)$ for all $i$ and $j$. For the SCC (nodes $B$, $C$ and $D$) we used multivariate Gaussian distribution with same means and variances (equal to 5) and pairwise correlations of 0.9. This way, all of generated data was very highly correlated, making it difficult to distinguish between different graphs using likelihood alone, even under the correct assumption about data generating mechanism being a~Gaussian linear additive noise model.

\begin{figure}[ht]
\centering
\includegraphics[width=1\textwidth]{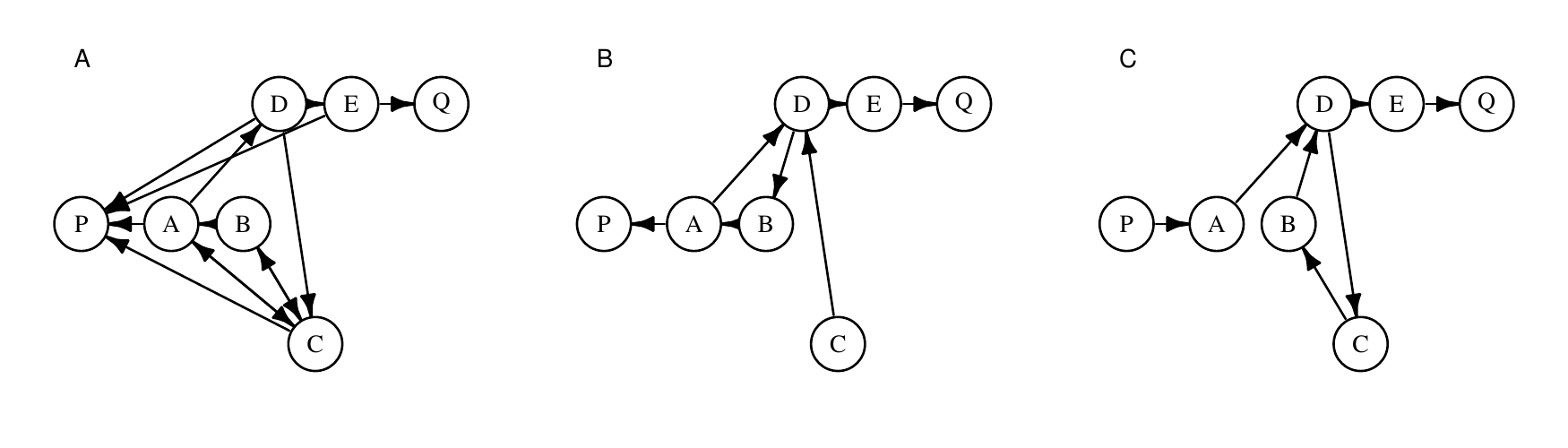}

\caption{Graph structure inference when likelihood is the same as the generative model. Highest scoring graphs for: (A) A model without informative priors. (B) A model with prior on out-degree and SCC size. (C) A model with additional prior on the ``first cause'' (no edges allowed into node $P$).}

\label{figure:case_linear}
\end{figure}

First, in the absence of prior information (Figure \ref{figure:case_linear}A) we did not succeed in retrieving the data generating graph and many superfluous edges were found. (Although we usually prefer to work with edge probabilities, for clarity of presentation we only show the best scoring graph here.) Including an~informative prior on the out-degree (power law with $\gamma = 3$) and size of SCCs (no larger than 3) enabled us to detect the SCC and the undirected edges correctly (Figure \ref{figure:case_linear}B). Lastly, adding information on the first cause, \textit{i.e.}, enforcing (through a~Bernoulli prior) $Pa(P) = \emptyset$, allowed us to retrieve the data generating graph (Figure \ref{figure:case_linear}C).

The last two steps illustrate two difficulties with learning network structures. First comes the problem of detecting dependencies from data, which can be helped by putting a~strong prior on the types of structures expected to occur in the graph (in this case degrees and SCC sizes). Even if we succeed in this, we are still left with multiple candidate graphs: in this case the condensed graph is a~path from $P$ to $Q$ which is Markov-equivalent to a~path from $Q$ to $P$. Only the addition of a~prior on whether $P$ or $Q$ is a~probable cause can help us retrieve the true network.

As discussed, the choice between SCCs and DAGs is highly sensitive to correlation. We repeated the simulation using the same data-generating mechanism, but with a~pairwise correlation of the SCC nodes equal to 0.5 instead of 0.9, DAGs were then preferred when using uninformative structural priors and the best graph resembled Figure \ref{figure:case_linear}(C) but without an~SCC. Setting $q = m + 11$ and scaling the ``default'' $\kappa$ appropriately is enough to bring up an~SCC again.

\subsection{SCCs detection in a~50-node linear model}

\begin{figure}[h]
\centering
\includegraphics[width=\textwidth]{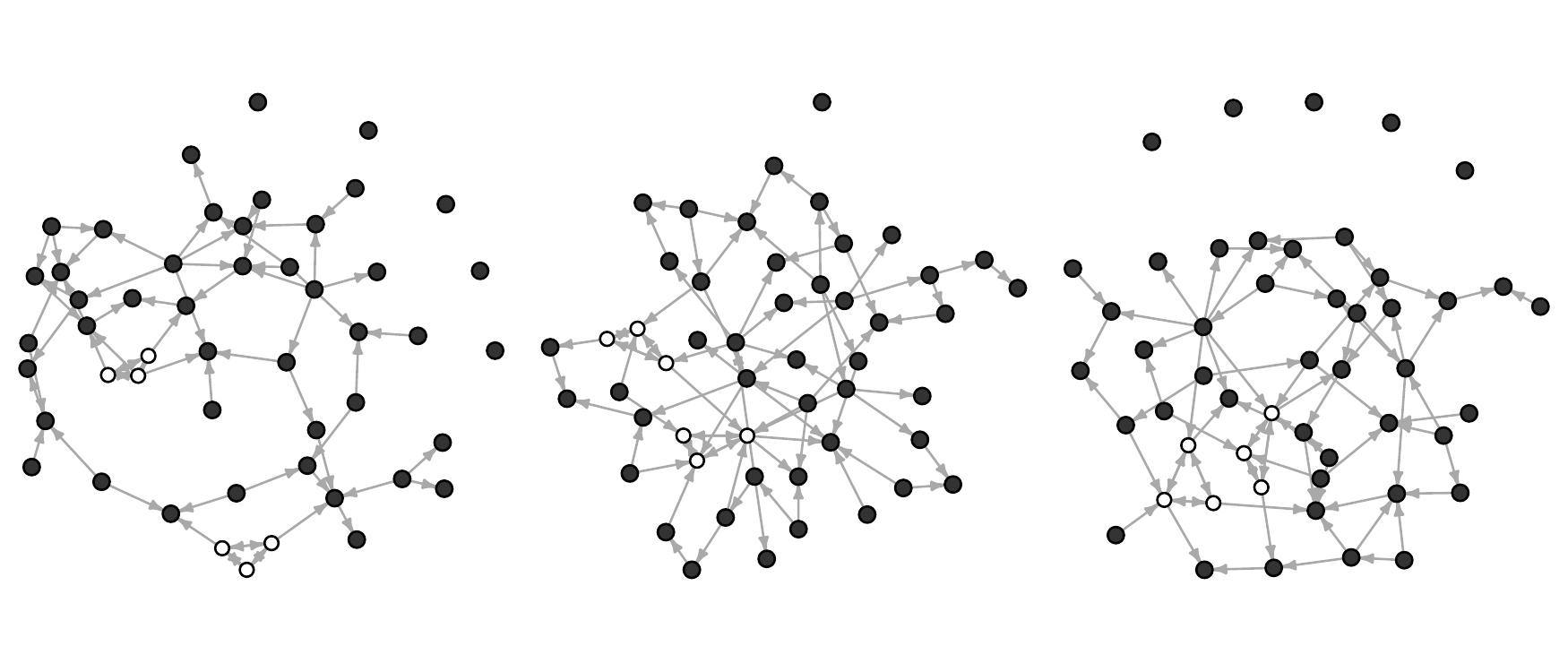}
\caption{Examples of three graphs (out of 800) from which data was generated for this performance benchmark. All graphs were randomly generated with the same settings. White nodes belong to (three-node) SCCs.}
\label{fig:case_large_graph}
\end{figure}

In the second simulated study, we generated batches of 100 DAGs of 50 nodes by randomly permuting nodes and drawing each edge $e_{ij}$ with probability of occurrence at 5\% if $j > i$ (to avoid SCCs). We then created two three-node SCCs in each graph by adding all possible edges between two groups of three randomly selected nodes. Example of such graphs are presented in Figure \ref{fig:case_large_graph}. Each graph was then used as~data generation mechanism for 100 data values for each node, according to a~normal linear model with regression coefficients set to 1. That is, for $j$-th node, the $i$-th generated value $x_j(i) = \sum_{X_k \in Pa(X_j)} x_k(i) + \epsilon_{ij}$, with $\epsilon_{ij} \sim \mathcal{N}(0, 1)$ i.i.d. for all $i$ and $j$. For SCCs the distribution was multivariate normal, with correlation between any two nodes fixed at 0.5 or 0.9, to benchmark performance in two different cases.

For inference, we compared four prior assumptions on the inverse Wishart parameters by varying $q$ and setting $\kappa$ to the desired correlation and scaling it appropriately (see Equation \ref{equation:expected_value}). We selected a~``default'' uninformative prior $q = m+1$ assuming a~within-SCC correlation of 0.5; a~$q = m+11$ prior (equivalent to 10 data points) assuming a~correlation of 0.5; a~$m+11$ prior assuming a~correlation of 0.9; and a~$m+101$ prior (equivalent to 100 data points) assuming a~correlation of 0.5. A summary of these combinations is given together with results in Table \ref{tab:case_large}. 

In each case, we set a~Bernoulli prior on the probability of edge occurrence $p(e_{ij}) = 0.025$ when $i \neq j$ (instead of 5\%, as in graph generation half of off-diagonal edges were not allowed) and constrained the size of SCCs to be at most three, but imposed no more priors. 

For each combination of ``true'' correlation and prior assumptions we generated data and ran MCMC inference 100 times. For each graph, we used 20 millions MCMC iterations, discarding first ten million, to infer on its structure. We only assessed MCMC convergence on a few selected graphs, but assumed that such run length was adequate given the simple nature of the problem and that we are interested in relative, not absolute, performance. For each of these runs, the probability of occurrence of SCCs was calculated from a~sample of 100 adjacency matrices drawn from the MCMC chain. We report the area under the receiver operating characteristic curve (AUROC). It is the same as described in Marbach et al.,  \cite{marbach_revealing_2010}, -- briefly, the $k$ possible edges in the graph were ordered by probabilities obtained from MCMC and we calculated sensitivity and specificity $k$ times, assuming that $1, 2, \ldots, k$ first edges occur and the rest do not. Note that perfect prediction (AUROC = 1) is impossible in this example, as we calculate our score for directed graphs and not equivalence classes. For SCCs, we only assessed sensitivity (as the AUROC statistic captures overall specificity well), by calculating a~probability that the ``true'' SCCs are present in the MCMC results. Under our definition we needed to ``detect'' all three nodes of the SCC to count as a~success. Table \ref{tab:case_large} presents results for both AUROC and ``SCC sensitivity''; results are averaged over 100 inferences for each row.

% latex table generated in R 3.5.0 by xtable 1.8-2 package
% Wed Jun 27 19:59:34 2018

\begin{table*}[h]
\caption{Sensitivity and specificity of the scoring method in retrieving true network structure.} 
% \caption{Sensitivity-specificity of the scoring method in detecting true causal graphs , and performance in detecting three-node SCCs  under different generative within-SCC correlations (``true Correlation'') and hyperparameter prior assumptions. All results are averaged over 100 simulated graphs.} 
\centering
\begin{tabular}{llcc}
  \hline
  True Correlation & Prior & AUROC & SCC Pr \\ 
  \hline
  0.50 & q = m + 1; Cor = 0.5     & 0.93 & 0.12 \\ 
  0.50 & q = m + 11; Cor = 0.5   & 0.91 & 0.26 \\ 
  0.50 & q = m + 11; Cor = 0.9   & 0.86 & 0.01 \\ 
  0.50 & q = m + 101; Cor = 0.5 & 0.91 & 0.52 \\ 
  0.90 & q = m + 1; Cor = 0.5     & 0.89 & 0.10 \\ 
  0.90 & q = m + 11; Cor = 0.5   & 0.89 & 0.40 \\ 
  0.90 & q = m + 11; Cor = 0.9   & 0.87 & 0.95 \\ 
  0.90 & q = m + 101; Cor = 0.5 & 0.91 & 0.39 \\ 
   \hline
\end{tabular}
\label{tab:case_large}
\end{table*}

\begin{figure}[!ht]
\centering
\includegraphics[width=\textwidth]{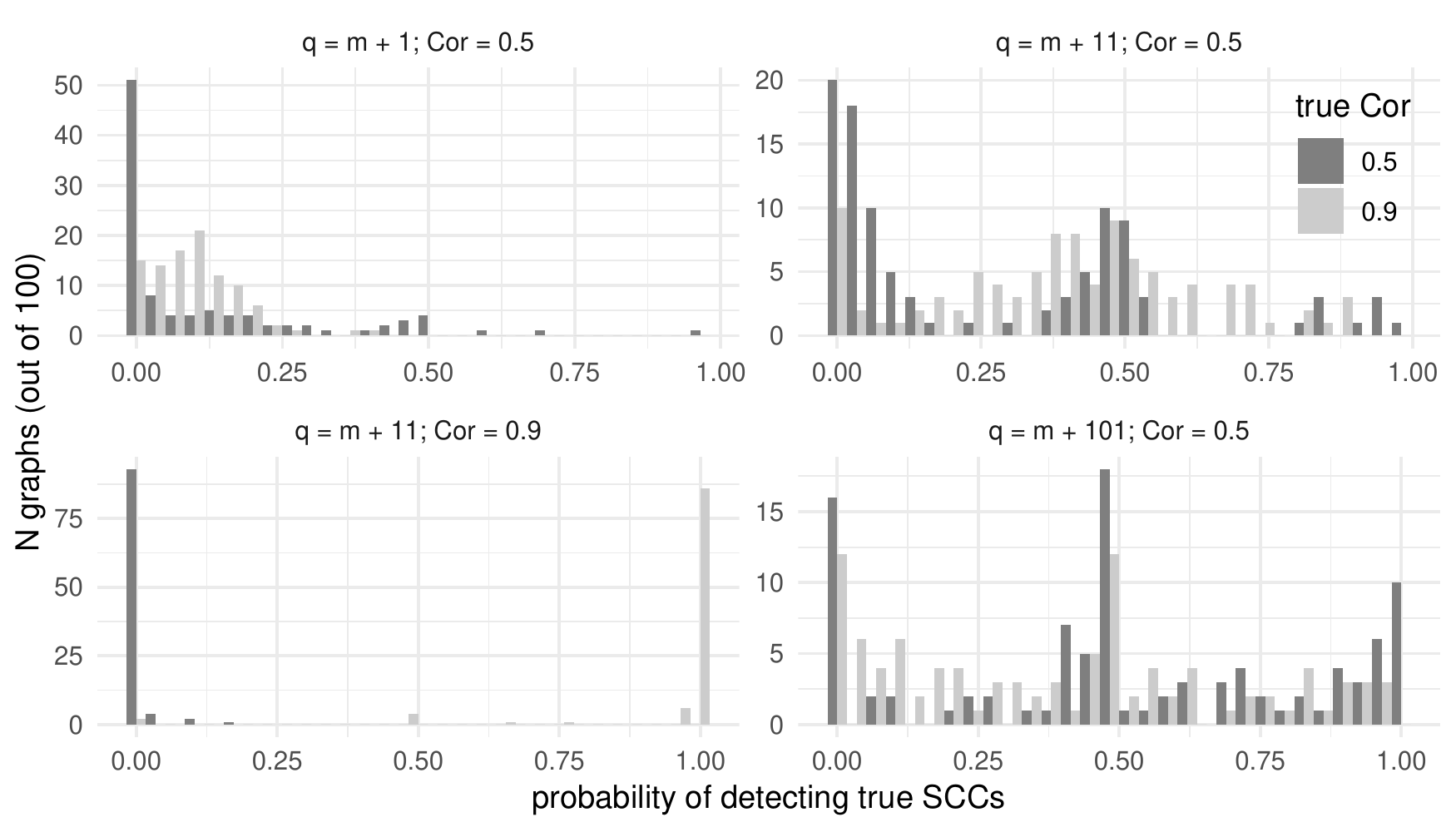}
\caption{How the probability of detecting true SCCs changes with ``true'' correlation (differently coloured histogram bars) and four priors (panels of the figure). Peaks at 0.5 correspond to cases where only one of two SCCs has been detected.}
\label{fig:case_large_distribution}
\end{figure}

Generally, the sensitivity and specificity (AUROC) of the score-based method is good under this simple generative model. However, the detection of SCCs is low with ``default'' settings, with about 10\% success rate. (Note that given that the equivalence class for three-node SCC is of size seven, \textit{i.e.} SCC and six DAG configurations, so we would expect success rate of about 14\% assuming equivalence of scores within class.) We can improve this by introducing informative priors. Generally highly correlated SCCs are easier to detect successfully, but using an~informative prior helps. Misspecification of prior does not seem to overly impact the overall (AUROC) performance, but does affect the detection of SCCs. That is most salient in the case of a~high correlation prior, as illustrated in the bottom left panel of Figure \ref{fig:case_large_distribution}. When assuming a~correlation of 0.50, the variability in success rate across graphs is large (with a~peak around probability of 50\%, corresponding to cases where one SCC has been identified perfectly and the other one not at all), but with a~prior on correlation equal to 0.90 the behaviour is completely different.

\section{DISCUSSION AND CONCLUSION}

%\textcolor{red}{[WW will rework this part] Ghislaine: I think in this part we should be more informative of %what we have done, i.e., this way of modelling is new   and suitable because \\ 
%\begin{itemize}
%\item 1) appropriate to graphs with  both cyclic and acyclic subgraphs  \\
%\item 2) can be seen  as an auxiliary modelling to improve the MCMC convergence  when exploring DAGs space \\
%\item 3) when undirecting the edges involved in cyclic parts,  the resulting graph  belongs to the class of %reciprocal graphs and hence benefits from the results on Markov property derived by Koster (96). In particular %the joint distribution on the nodes set we have considered is consistent with the Markov property of reciproca%l graphs. \\ 
%\item 4) to our knowledge this is the first attempt to modelling such a graph with a Bayesian approach. To %some extend, this work generalizes the ones of Datta et al (17)  and Mukherjee and Speed (08)  to general %directed graphs. \\
%\item 5) Then we can move to the computational part.  \\
%\item 6) We can conclude by "In this paper, we focus on studying deeply the behavior of our modelling w.r.t. %several quantities : hyperparameters, informative/uninformative struture priors... 
%\end{itemize}}

We proposed a~model to represent cyclic structures within Bayesian networks. Our model offers an~alternative way of describing joint probability distribution and performing network inference without apparent computational drawbacks.
In our approach, SCCs are condensed to form multivariate nodes, which are still embedded in an~acyclic Bayesian network. We can therefore factorise the likelihood, a~key computational advantage of Bayesian networks. 
We use a~score-based approach in a~fully Bayesian setting. A posterior sample of graphs is obtained by MCMC sampling. This allows us to integrate prior knowledge on presence of edges, degrees, acyclic motifs, occurrence of SCCs \textit{etc}. The placement of informative priors on network structure also brings faster convergence of MCMC sampling (if the data are not conflicting with the prior) by putting soft constraints of the size of the set of likely graphs. 

The likelihood model we present is an additive linear model with Gaussian noise. Such model allows us to easily compute score by integrating out parameters. The only (arbitrary) constraint imposed by our Gaussian model is that the number of parents for all members of an~SCC has to be less than the number of data points for each node. In the future other models for likelihood or other scoring functions should be explored. We also note that in the present form the model can only account for time as an~additional linear term in regression, although a~dynamic version of it might be workable. Use of SCCs with discrete random variables should also be explored.

Many alternative methods for characterising dependencies in graphs containing cycles have been proposed, including reciprocal graph models based on work by  Koster \cite{koster_markov_1996} (see  paper by Ni and al.,  \cite{ni_reciprocal_2018}, for recent application)  or a heuristic algorithm approach to learning cycles from experimental data by Itany et al., \cite{itani_structure_2010}. Our work differs from statistical models for purpose of learning causal relationships in observational data, as under our model Markov-equivalent graphs can have different scores due to choice of priors and hyperpriors relating to SCCs. 

Under the proposed model, detection of SCCs is sensitive to the choice of hyperparameters. Informative priors can be used to promote or suppress occurence of SCCs in the posterior. We can choose priors to favour SCCs over DAG structures even when correlation is lower than the threshold visible in Figure \ref{figure:lvg}. This may be useful in applications where we know \textit{a priori} that cycles are present or simply wish to describe joint distribution differently. However, in our model, a~limitation in setting informative priors is tied to the properties of $\mathcal{W}^{-1}$ distribution, where the precision of variances and correlations is governed by a~single parameter, $q$. Here again we must decide between standardisation and flexibility. 

Our simulation study with a~small linear model also shows the importance of prior choices on the inference. Note that pure likelihood-based inference amounts to placing only an~indifferent Bernoulli prior on the adjacency matrix, and would bring the same inefficient inference as in Figure \ref{figure:case_linear}B. In the case of a~larger network, the inference scales up well, but SCCs typically have a~50\% chance to be detected. 

Finally, besides substantive applications, allowing for SCCs can reduce computation time (by reducing the number of nodes) even for underlying DAGs, and improves convergence by easing edge reversals. The practical impact of those computational benefits should be explored in greater detail in the future.

%% ** The bibliograhy **
%\bibliographystyle{ba}

% ** Acknowledgements **
%\acknowledgements{

\smallskip
\noindent \textbf{Acknowledgments}
F. Bois' work was funded by the Horizon 2020 project "EU-ToxRisk" of the European Commission  (Contract 681002).

\newpage

\section*{Supplementary materials for ``Structure learning of Bayesian networks involving cyclic structures''}

\noindent We start by restating the main equation of the Section 2 in the main paper. The joint posterior distribution of $(\theta, \Sigma,Y)$ conditional on $X$ can be expressed as

\begin{equation}
\label{model_posterior_1} 
P(\theta, \Sigma, Y \vert  X) =   P(\theta \vert \Sigma, Y,X) \cdot P(\Sigma \vert Y,X) \cdot  P(Y\vert X), 
\end{equation}

\noindent where  $P(Y\vert X)$,  $P(\theta \vert \Sigma, Y,X)$ and  $P(\Sigma \vert Y,X)$  are of the form:

\begin{align}\label{pre} 
P(Y\vert X)  = & (2 \pi)^{\frac{m(k-n)}{2}} 2^{\frac{m(p-q)}{2}} \dfrac{\Gamma_{m}(\frac{p}{2}) }{ \Gamma_{m}(\frac{q}{2})} 
\vert\kappa \vert ^{\frac{q}{2}}  \vert X^t X \vert ^{-\frac{m}{2}} \vert \kappa + A_0 \vert ^{-\frac{p}{2}},\\
% C_0 &=& \dfrac{\vert \kappa \vert ^{\frac{q}{2}}}{2 ^{\frac{mq}{2}} \Gamma_{m}(\frac{q}{2})} \dfrac{1}{(2 \pi)^{nm/2}} (2 \pi)^{mk/2} \vert X^t X \vert ^{\frac{-m}{2}} \vert \kappa + A_0 \vert ^{-\frac{p}{2}} 2^{\frac{mp}{2}} \Gamma_{m}(\frac{p}{2}), \label{pre}\\
P(\theta \vert \Sigma, Y,X) = & \dfrac{\vert X^t X \vert ^{\frac{m}{2}} \vert \Sigma \vert^{-k/2}}{(2\pi) ^{\frac{mk}{2}}}
\exp \left\lbrace -\dfrac{1}{2} tr\Sigma^{-1} (\theta - \hat{\theta})^t X^t X(\theta - \hat{\theta})\right\rbrace, \nonumber\\
P(\Sigma \vert Y,X) = & \dfrac{\vert \Sigma \vert^{-(p+m+1)/2}}{2 ^{\frac{mp}{2}}\Gamma_{m}(\frac{p}{2})}
\exp \left\lbrace - \dfrac{1}{2} tr \Sigma ^{-1}(\kappa + A_0) \right\rbrace \vert \kappa + A_0 \vert ^{\frac{p}{2}},\nonumber
\end{align}

\noindent where $p = q+n-k$. 

\noindent When $\theta$ and $\Sigma$ are integrated out from (\ref{model_posterior_1}), we obtain (\ref{pre}), which corresponds to the marginal likelihood function of the strongly connected component under the \textit{constant-Wishart} model, that is, 

$$P(Y\vert X)= s(\mathbf{{x_j}^L}\vert Pa(\mathbf{{x_j}^L})).$$

\section*{S1: EQUIVALENCE OF INVERSE GAMMA AND INVERSE WISHART MARGINAL LIKELIHOODS}\label{supplement:equivalence}

We will show that the marginal likelihood obtained with the inverse Wishart distribution with $m=1$ coincides with the inverse Gamma case. 
We use the same notation as in Section 2 of the paper.
%\ref{section:data_likelihood}. 
Recall that we model the distribution of $x_j \vert Pa (x_j)$ using the linear regression model with $m=1$, $X$ the design matrix and $Y=x_j$ and denote by $\theta$ and $\sigma^2$ the model parameters, where $\theta$ is a~$k$-vector and $\sigma^2$ is a~positive real number. 
Moreover, recall that the least square estimate ${\hat \theta}$ and the sample covariance matrix $A_0$ are given by: 
 ${\hat \theta}=(X^t X)^{-1} X^t Y$ and $A_0=(Y-X{\hat \theta})^t (Y-X{\hat \theta})$. 
 
Then the likelihood is 
$$ L(\theta,\sigma^2; Y,X)=\frac{1}{(2 \pi \sigma^2)^{n/2}}\exp(-\frac{1}{2\sigma^2} (Y-X\theta)^t (Y-X\theta)).$$

Let us consider the following independent priors on $\theta$ and $\sigma^2$:

$$\theta \propto Constant, \quad \quad \sigma^2 \sim Inv Gamma (q/2, \tau/2).$$ 

The marginal likelihood is then obtained from 
\begin{align*}
s(Y|Pa(Y)) =& \int \displaystyle{\frac{\tau^{q/2}}{2^{q/2} \Gamma(q/2)(\sigma^2)^{q/2 +1}}} \\
 & \exp(-\displaystyle{\frac{\tau}{2\sigma^2}}) 
 \left(\int L(\theta,\sigma^2; Y,X) d\theta\right) d\sigma^2.
\end{align*}

First, integrating out $\theta$ leads to

\begin{align*}
\int L(\theta,\sigma^2; Y,X) d\theta
%& =& 
%\frac{1}{(2 \pi \sigma^2)^{n/2}}\exp(-\frac{1}{2\sigma^2} 
 %(Y-X{\hat \theta})^t (x-M{\hat \theta})) \\ 
% && \times (2\pi\sigma^2)^{(k)/2} |X^t X|^{-1/2} \\
% & & \times \int \frac{|X^t X|^{1/2}}{(2\pi\sigma^2)^{k/2}} \exp(-\frac{1}{2\sigma^2} 
% ({\hat \beta}-\beta)^t M^t M ({\hat \theta}-\theta)) d\\theta\\
 =& \displaystyle{\frac{(2\pi\sigma^2)^{k/2} \; |X^t X|^{-1/2}}{(2 \pi \sigma^2)^{n/2}}} \\
 &\exp\left(-\displaystyle{\frac{1}{2\sigma^2} }
 (Y-X{\hat \theta})^t (Y-X{\hat \theta})\right) 
\end{align*}

Second, integrating out $\sigma^2$ leads to $s(Y|Pa(Y))$; indeed,

\begin{eqnarray}\label{prior_predict}
s(Y|Pa(Y))
%& =& \int \displaystyle{\frac{\tau^{q/2}}{2^{q/2} \Gamma(q/2)(\sigma^2)^{q/2 +1}}} \exp(-\displaystyle{\frac{\tau}{2\sigma^2}}) \displaystyle{\frac{(2\pi\sigma^2)^{k/2} \; |X^t X|^{-1/2}}{(2 \pi \sigma^2)^{n/2}}} \exp\left(-\displaystyle{\frac{A_0}{2\sigma^2} }\right) d\sigma^2 \nonumber \\
 %& =& \displaystyle{\frac{\tau^{q/2}(2\pi)^{k/2} |X^t X|^{-1/2}}{2^{q/2} (2 \pi)^{n/2} \Gamma(q/2)}} \int \displaystyle{\frac{1}{(\sigma^2)^{n/2+q/2 -k/2 +1}}} 
 %\exp\left(-\displaystyle{\frac{1}{2\sigma^2}(\tau+ A_0)}\right) d\sigma^2 \nonumber \\
&=& \displaystyle{\frac{\tau^{q/2}(2\pi)^{k/2} |X^t X|^{-1/2}}{2^{q/2} (2 \pi)^{n/2} \Gamma(q/2)}} \times 
 \displaystyle{\frac{ 2^{\frac{n+q -k}{2}}\Gamma(\frac{n+q -k}{2})}{(\tau+ A_0)^{\frac{n+q -k}{2}}}} \nonumber \\
&& \times \int \displaystyle{\frac{(\tau+ A_0)^{\frac{n+q -k}{2}}}{ 2^{\frac{n+q -k}{2}}\Gamma(\frac{n+q -k}{2})}} \displaystyle{\frac{1}{(\sigma^2)^{\frac{n+q -k}{2} +1}}} \nonumber\\
&& \exp\left(-\displaystyle{\frac{1}{2\sigma^2}(\tau+ A_0)}\right) d\sigma^2 \nonumber \\
&=& (2\pi)^{\frac{k-n}{2}} \; 2^{\frac{n -k}{2}} \; \displaystyle{ \frac{ \Gamma(\frac{n+q -k}{2}) }{ \Gamma(q/2)} } \tau^{q/2} \; |X^t X|^{-1/2} \; \nonumber \\
&& (\tau+ A_0)^{-\frac{n+q -k}{2}}, 
\end{eqnarray}
which corresponds to the marginal likelihood for the SCC case defined in Section 2 of the paper,
%Section \ref{section:data_likelihood}, 
assuming $m=1$.

\section*{S2: MARGINAL LIKELIHOOD FOR TWO-NODE GRAPHS}
\label{supplement:loop_dag}
We will now show that the marginal likelihood's for SCCs and full graphs are not equal under our model, on an example of two-node graphs. We denote the nodes by
$A$ and $B$ and focus on marginal likelihood's for an~empty graph $\mathcal{G}_1$, SCC graph $\mathcal{G}_2$ ($A\leftrightarrow B$) and DAG $B \rightarrow A$, $\mathcal{G}_3$. All are evaluated under the same data $D=(x_A,x_B)$. 
 
% Notation: $1_n=\begin{pmatrix} 1 \\ 1 \\ \ldots \\ 1\end{pmatrix}$ is the vector of ones of dim $n$, 
We denote by $1_n$ a~column vector of ones of dim $n$, ${\bar x}_{A}$ and ${\bar x}_{B}$ are the empirical mean of $x_A$ and $x_B$ respectively while $s_A^2$ and $s_B^2$ are the empirical variances of $x_A$ and $x_B$; $s_{A,B}$ denotes the empirical covariance between $x_A$ and $x_B$ and $\sigma_{A,B}$ denotes the empirical variance covariance matrix of $D$.

\medskip\noindent
\textit{Empty graph:} The marginal likelihood for the empty graph is equal to $s(x_A \vert Pa(x_A) ) \times s(x_B\vert Pa(x_B) )$, where $ Pa(x_A) = Pa(x_B) = \emptyset$ and each term is given by Equation (\ref{prior_predict}) with $X=1_n$, $|X^t X|=n$, $A_{0,B}=(x_B - 1_n {\bar x}_B)^t (x_B - 1_n {\bar x}_B)$ and $A_{0,A}=(x_A - 1_n {\bar x}_A)^t (x_A - 1_n {\bar x}_A)$; we then have, 

\begin{align*}
s(x_A,x_B \vert \mathcal{G}_1) 
=& (2\pi)^{1-n} \; 2^{n -1} \; \displaystyle{ \left(\frac{ \Gamma(\frac{n+q -1}{2}) }{ \Gamma(\frac{q }{2})} \right)^2} \frac{\tau^{q}}{n} \; \\
& [(\tau+ ns_B^2)(\tau+ ns_A^2)]^{-\frac{n+q -1}{2}}.
 \end{align*}

%&=&s(x_A \vert Pa(x_A) ) \times s(x_B\vert Pa(x_B) ) \\ 

%& =& \displaystyle{\frac{\tau^{q/2}}{{\sqrt n}(\tau+ (x_A - 1_n {\bar x}_A)^t (x_A - 1_n {\bar x}_A))^{\frac{n+q -1}{2}}}} \; %\displaystyle{\frac{2^{\frac{n -1}{2}}}{ (2 \pi)^{n/2- 1/2}}}\; \displaystyle{\frac{\Gamma(\frac{n+q -1}{2}) }{ \Gamma(q/2)}} \\ 
%& \times & \displaystyle{\frac{\tau^{q/2}}{{\sqrt n}(\tau+ (x_B - 1_n {\bar x}_B)^t (x_B - 1_n {\bar x}_B))^{\frac{n+q -1}{2}}}} %\; \displaystyle{\frac{2^{\frac{n -1}{2}}}{ (2 \pi)^{n/2- 1/2}}}\; \displaystyle{\frac{\Gamma(\frac{n+q -1}{2}) }{ \Gamma(q/2)}} \

\medskip\noindent
\textit{SCC graph:} The marginal likelihood for $\mathcal{G}_2$ is given by $C_0$ defined in \eqref{pre} with $m=2$, $k=1$ since 
$$Pa (x_A,x_b)=\emptyset$$ and 
$$A_0=\begin{pmatrix} x_A -{\bar x}_A 1_n & x_B -{\bar x}_B 1_n\end{pmatrix}^t \begin{pmatrix} x_A -{\bar x}_A 1_n & x_B -{\bar x}_B 1_n\end{pmatrix}$$, that is $A_0 = n \sigma_{A,B}.$ 

Finally, the marginal likelihood is 

\begin{align*}
s(x_A,x_B| \mathcal{G}_2) =& (2 \pi)^{-\frac{2(n-1)}{2}} 2^{\frac{2(n-1)}{2}} \dfrac{\Gamma_{2}(\frac{n+q-1}{2}) }{ \Gamma_{2}(\frac{q}{2})} \\
& \vert \kappa \vert^{\frac{q}{2}} \vert n\vert ^{-\frac{2}{2}} \vert \kappa + A_0 \vert ^{-\frac{n+q-1}{2}}.\end{align*}

\medskip\noindent
\textit{DAG graph:} 
\noindent The marginal likelihood for $\mathcal{G}_3$ is defined by $s(D\vert \mathcal{G}_3)= s( x_A \vert x_B) \times s(x_B\vert Pa(x_B))$ where 
$Pa(x_B) =\emptyset$.

\noindent Likelihood $s(x_B\vert Pa(x_B))$ is given by Equation (\ref{prior_predict}) with $X=1_n$, $|X^t X|=n$, and $A_{0,B}=(x_B - 1_n {\bar x}_B)^t (x_B - 1_n {\bar x}_B)$, while $s( x_A \vert x_B)$ is defined with
\begin{align*}
k &= 2, \\
X &= \begin{pmatrix} 1 & 1 & \ldots & 1 \\ x_{B,1} & x_{B,2} & \ldots & x_{B,n} \end{pmatrix}^t , \\
X^t X &=\begin{pmatrix} n & n{\bar x}_{B} \\ n{\bar x}_{B} & \sum_{i}x_{B,i}^2 \end{pmatrix}, \\
X{\hat \theta} &= \begin{pmatrix} \left(( {\bar x}_A + (x_{B,i} - {\bar x}_B) \displaystyle{\frac{s_{A,B}}{ s_B^2}}) \right)_i \end{pmatrix}, \\
A_0 &= ns_A^2 - n(s_{A,B}^2 / s_B^2). 
\end{align*}

%\begin{eqnarray*}
%A_0 % &=& \left(x_A - X{\hat \beta}\right)^t \left(x_A - X{\hat \beta}\right)\\ 
%& =& \left(x_A - {\bar x}_A1_n \right)^t \left(x_A -{\bar x}_A1_n\right) -n\displaystyle{\frac{s_{A,B}^2}{s_B^2}}\\
%& =& ns_A^2 -n\displaystyle{\frac{s_{A,B}^2}{s_B^2}}
 %\end{eqnarray*}
 
\noindent
Combining the above terms provides the marginal likelihood for ${\mathcal G}_3$: 

\begin{align*}
s(x_A,x_B| \mathcal{G}_3) =& 
 (2 \pi)^{-\frac{2n-3}{2}} 2^{\frac{2n-3}{2}} \dfrac{\Gamma(\frac{n+q-1}{2})\Gamma(\frac{n+q-2}{2}) }
 { (\Gamma (\frac{q}{2}))^2} \tau^{q} \\ 
& \frac{ \vert \tau + ns_B^2 \vert ^{-\frac{n+q-1}{2}}}{n^{1/2}}
 \frac{ \vert \tau+ ns_A^2 - n\frac{s_{A,B}^2}{s_B^2} \vert ^{-\frac{n+q-2}{2}}}{(ns_B^2)^{1/2}}. \nonumber
\end{align*}

This expression depends on $s_B^2$ and $s_A^2$ in a~way that does not allow for equivalence between $A\rightarrow B$ and $B\rightarrow A$.

\section*{S3. ALGORITHM FOR UPDATING LIKELIHOOD IN SCC CASES}\label{supplement:flowchart}

We present here a simplified flowchart for updating the graph score when edges are added or deleted with minimum of necessary re-calculations. By ``updating'' we mean calculating difference(s) in $l$, logarithm of node's contribution to marginalised likelihood, between steps, which then allows us to evaluate if the jumping proposal is accepted (and if yes, to update the score function). We denote by $L$ sum of $l$'s over all nodes.

In a DAG case only the child's conditional probability is affected when we change edges. Therefore from step to step we only need to store a vector of $l$ for individual nodes, a proposed calue of $l$ for child, and the current value of $L$ (sum of all scores)\footnote{In practice we also store and dynamically update parent sets and their sizes for each node, as this information is used every time score is evaluated.}. If proposal to add or remove an edge is accepted, we update the $L$ by a difference in child values before and after.

In a cyclic graph the incremental update of $L$ is more difficult. We need to store additional structure describing SCCs: their sizes and member nodes for each. An addition of an edge may result in ``closing a loop'' and deletion may ``destroy'' an SCC. Moreover, in case of addition we may be creating one larger SCCs out of two smaller SCCs while in the case of removal a new, smaller SCCs may appear where a bigger one was deleted. In the worst case, $l$ values in all nodes can change as a consequence of a single addition or removal. We avoid re-calculating for all nodes where possible by using a set of simple \textit{if-else} rules outlined below. 

%In graph $\mathcal{G}$, for addition/deletion from node $i$ to node $j$.

% Define block styles
\tikzstyle{decision} = [diamond, draw, fill=gray!20, 
    text width=4.5em, text badly centered, node distance=3cm, inner sep=0pt]
\tikzstyle{block} = [rectangle, draw, fill=blue!20, 
    text width=8em, text centered, rounded corners, minimum height=4em]
\tikzstyle{line}  = [draw, -latex']
\tikzstyle{cloud} = [draw, ellipse,fill=red!20, node distance=3cm,
    minimum height=2em]
    
\begin{tikzpicture}[node distance = 4cm, auto]
    % Place nodes
    \node [decision] (init) {addition or deletion?};
    \node [decision, below left of=init]  (add1) {parent \& child in same SCC?};
    \node [cloud,    left  of=add1] (add11){End};
    \node [block,    below of=add1] (add2)  {Update $SCC(\mathcal{G})$ (Tarjan)};
    \node [block,    below of=add2] (add21) {update child's $l$};
    \node [decision, below of=add21] (add3) {parent \& child in same SCC?};
    \node [cloud,    left  of=add3] (add31){End};
    \node [block,    below of=add3] (add4) {subtract $l$ of nodes that ``got in'' the SCC};
    
    \node [decision, below right of=init]  (del1) {child \& parent in the same SCC?};
    \node [block,    below of=del1]  (del2) {Update $SCC(\mathcal{G})$ (Tarjan)};
    \node [decision, below of=del2]  (del3) {child's SCC now smaller?};
    \node [block,    right of=del3]  (del31){update child's $l$ (lost parent)};
    \node [block,    below of=del3]  (del4) {recalculate $l$ for all members of ``old'' SCC};
    
    \path [line] (init) -| node {add} (add1);
    \path [line] (add1) -- node  {no} (add2)  ;
    \path [line] (add1) -- node {yes} (add11) ;
    \path [line] (add2) --              (add21);
    \path [line] (add21)--              (add3);
    \path [line] (add3) -- node  {no} (add31) ;
    \path [line] (add3) -- node {yes} (add4)  ;
    
    \path [line] (init) -| node {delete}(del1);
    \path [line] (del1) -- node {yes}   (del2);
    \path [line] (del1) -| node {no}   (del31);
    \path [line] (del2) --              (del3);
    \path [line] (del3) -- node  {no}   (del31);
    \path [line] (del3) -- node {yes}   (del4);
    
\end{tikzpicture}
\end{document}